\documentclass[journal]{IEEEtran}

\usepackage[extra]{tipa}
\usepackage{setspace}    
\usepackage{cite}
\usepackage{graphicx}
\usepackage{amssymb}
\usepackage{epstopdf}
\usepackage[low-sup]{subdepth}
\usepackage[utf8]{inputenc}
\usepackage{amssymb} 
\usepackage{amsmath}

\begin{document}

\title{Practical Full Duplex Physical Layer \\ Network Coding}

\author{
        Semiha Tedik, ~\IEEEmembership{Student Member,~IEEE}, and G\"{u}ne\c{s} Karabulut Kurt, ~\IEEEmembership{Member,~IEEE}
\thanks{S. Tedik and G. Karabulut Kurt are with the Department
of Electronics and Telecommunications Engineering, Istanbul Technical University, İstanbul, Turkey.
e-mail: \{tedik, gkurt\}@itu.edu.tr}
}


\maketitle

\begin{abstract}
We propose a practical network code for the wireless two-way relay channel where all nodes communicate in full duplex (FD) mode. The physical layer network coding (PNC) operation is applied with the FD operating nodes, reducing the transmission time to a single time slot, hence doubling the spectral efficiency when compared to classical PNC systems. In our system model,  binary phase shift keying modulated signals are transmitted over Rayleigh fading channels.  We derive the theoretical error rates at relay and end nodes according to the maximum likelihood detection rule,  in case of non-ideal self-interference cancellation. Theoretical results are also verified via simulations.
\IEEEpeerreviewmaketitle
\end{abstract}


\IEEEpeerreviewmaketitle

\section{Introduction}

\begin{figure}[tb]
 \centering
 \includegraphics[width=0.500\textwidth, keepaspectratio=true] {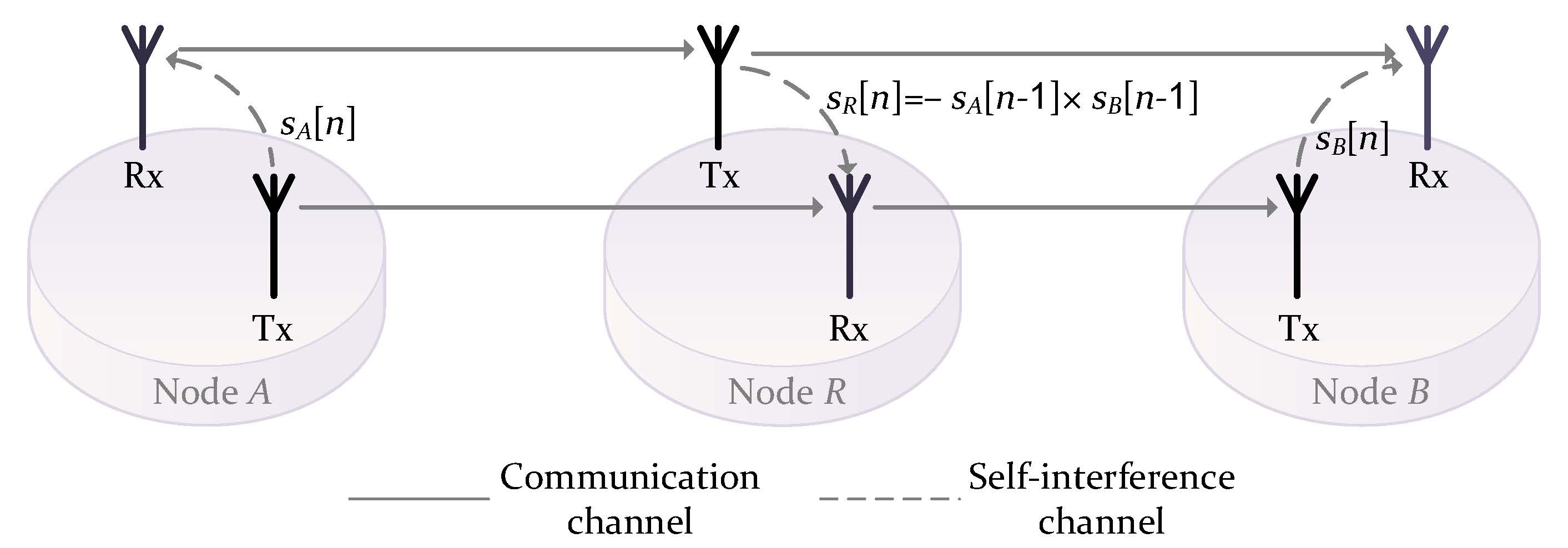}
 \caption{Full duplex physical layer network coding system. All nodes transmit in a single time slot. The relay transmits a function of the transmitted signals from end nodes via detect-and-forward relaying.}
 \label{fig:1}
\end{figure}

\IEEEPARstart{F}OLLOWING  the seminal works  \cite{ zang_PNC2} and \cite{ zang_PNC} there has been a surge of research activities in the field of physical layer network coding (PNC). The main principle of the PNC relies on the use of the two-way relay channel (TWRC) shown in Fig. \ref{fig:1} in a two phase process. The first phase is the multiple access (MA) phase, where end nodes $A$ and $B$ transmit their symbols to the relay node, $R$. In the second phase, referred to as the broadcast (BC) phase, $R$ transmits a function of the collided symbols from end nodes $A$ and $B$. Having the $a \; priori$ knowledge about their own transmitted symbols, $A$ and $B$ can detect the unknown symbol. Information exchange through  PNC hence takes two time slots in TWRC. 
 
In this paper, we propose a new physical layer network code for TWRC where the transmission takes a single time slot via combining the MA and BC phases by making use of full duplex (FD) transmission at all nodes. In the proposed network coding technique, named as full duplex physical layer network coding (FD-PNC), $R$ uses detect-and-forward (DF) relaying technique as shown in Fig. \ref{fig:1}. The maximum likelihood (ML) based joint detection is used at the relay in order  to eliminate the multiple access interference, which occurs due to the reception of two symbols from distinct sources at the same time slot. We adopt the max-log approximation to the ML detection rule, which is frequently preferred in order to reduce the computational complexity of the receiver \cite{max_log_ap}.  The proposed FD-PNC system can  save $50 \%$ of the time needed by the classical PNC approaches. Note that the FD-PNC saves $66.7 \%$ of the time when compared to the classical network coding, where the MA phase takes two time slots (requiring 3 time slots in total  \cite{network_coding_bir}). 

Despite the apparent advantages, making use of FD nodes has its challenges in terms of the generated self-interference (SI). In the literature, there are various SI cancellation techniques that are proposed in order to combat the associated performance degradation. Although the majority of the literature on FD transmission assume that the SI can ideally be eliminated \cite{PNC_FD,nam_PNC,tabataba_FD_PNC,TWRC_FD_relaying}, noting its deteriorating effects on the  performance, throughout our analyses we take the residual SI into account. In this paper, after describing the details of the system model we derive the corresponding bit error rate (BER) expressions for the relay node and  end nodes in the presence of Rayleigh fading channels for binary phase shift keying (BPSK) modulation. We verify our theoretical results via simulations.

\subsection{Related Work}
The use of FD nodes in TWRC is proposed for the first time in the pioneering work of  Rankov and  Wittneben, where authors investigate the information theoretical rate region without considering bit level transmission issues such as modulation and coding techniques \cite{PNC_FD}. A new lattice code is proposed for TWRC with FD nodes under non-fading channels, and the corresponding rate region analysis is performed in \cite{nam_PNC}. In \cite{tabataba_FD_PNC}, authors examine the effects of channel estimation errors on achievable rate region of the amplify-and-forward PNC system, again without including a bit level analysis.  New linear relaying functions are proposed to maximize the mutual information between source nodes in \cite{TWRC_FD_relaying}, again without considering physical layer transmission issues. The studies of TWRC with FD nodes listed above explore the information-theoretic standpoints. 
In the literature there is one recently proposed practical transmission scheme for relay communications with FD transmission only at the relay node \cite{one_way_FD}, where the single-way information transfer is completed in two time slots.   

In the literature about FD transmission, there are several proposed SI cancellation techniques such as as active  and passive  cancellation, operating at  RF and baseband \cite{mobicom_ilk, mobicom_son, duarte}. These works do not apply network coding. 

\section{System Model}
\label{ch:2}
\begin{figure*}[bt]
 \centering
 \includegraphics[width=1.00\textwidth, keepaspectratio=true] {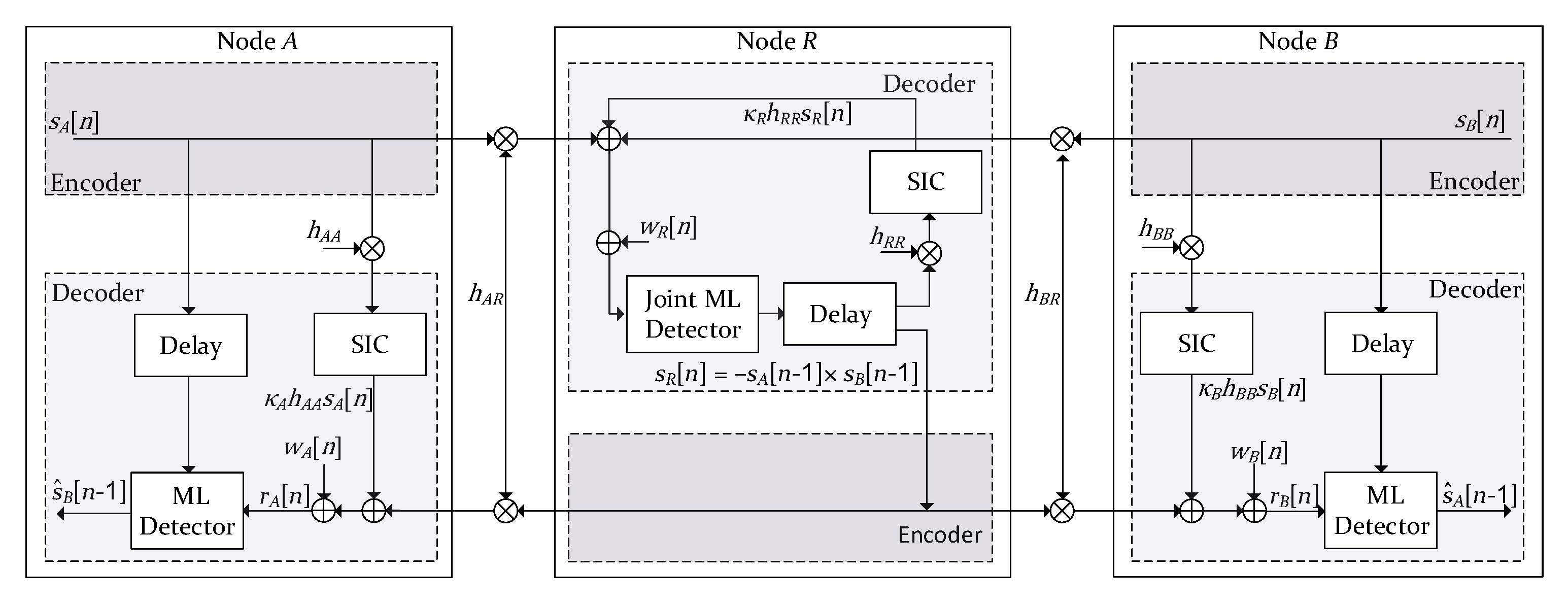}
 \caption{The block diagram of proposed model of full duplex physical layer network coding system (SIC: Self-interference cancellation, ML: Maximum likelihood). While nodes $A$ and $B$ transmit their information bits, the relay node transmits the modulated  XOR of detected bits from the end nodes. Hence two information bits are transmitted within a single time slot.}
 \label{fig:2}
\end{figure*}
The proposed system model consists of two end nodes
and one relay node indicated by $A$, $B$, and $R$, respectively. All nodes communicate in FD mode with separate transmit and receive antenna chains, as shown in Fig. \ref{fig:1}. There is no communication between the end nodes.  $d_i\in\{1,0\}$ denotes the  information bit of node $i$ and $s_i  \in\{1,-1\}$ represents the corresponding  BPSK modulated symbol, $i=A,B$. In this TWRC, we have two types of channels; the communication channel between the nodes, and the SI channel observed due to FD transmission. Let the communication channels be denoted by $h_{AR}$ and $h_{BR}$ for $A-R$ and $B-R$ links, respectively.   To simplify our notation we make use of the  frequently utilized channel reciprocity assumption, and state that $h_{RA}=h_{AR}$ and $h_{RB}=h_{BR}$ \cite{PNC_FD,zang_PNC,achievable_rate_HD}. The SI channels of $A$, $B$ and  $R$ are represented by $h_{AA}$, $h_{BB}$ and $h_{RR}$. Note that the time index for the channels are also omitted for notational simplicity. All channels are assumed to Rayleigh fading with unit energy. Block diagram of the system model is shown in Fig. \ref{fig:2}.

In the proposed system, the received signal at $R$ in the $n^{th}$ time slot can  be modeled as
\begin{eqnarray}
r_R[n] =  h_{AR}s_A[n]+  h_{BR}s_B[n]+ \kappa_{R}h_{RR}s_R[n] + w_R[n],      
\label{eq:1}
\end{eqnarray} 
where $w_R[n]$ denotes the additive white Gaussian noise (AWGN) with  the distribution $\mathcal{N}(0, \sigma_R^2)$. Here, $s_R[n]$ is the symbol that is transmitted by $R$ at the $n^{th}$ time slot. The term $\kappa_R$  represents the residual SI coefficient associated with FD transmission, and it is used to quantify the associated performance degradation (due to practical implementation constraints, SI  can not always be ideally canceled \cite{mobicom_ilk, mobicom_son}). 
 
Representing the average bit energy of the relay node by  $E_R$, the energy of the residual SI is  equal to $\kappa_R^2 E_R$. Noting the independence of the noise process and the SI, the average signal to interference and noise ratio (SINR) of the signal transmitted by node $i$ at relay node $R$ is \cite{proakis}
\begin{eqnarray}
\gamma_i = \dfrac{E \left[ \left|  (h_{iR} s_i[n])\right|^2 \right]}{E\left[ \left|  w_R[n] + (\kappa_{R}h_{RR}s_R[n]) \right|^2 \right]} = \dfrac{E_i}{\sigma_i^2 + \kappa_R^2 E_R }  
\label{eq:2}
\end{eqnarray}
where $E_i$ represents the average bit energy of the $i^{th}$ node. 
The SI term acts as the Gaussian interference  \cite{JD_1, Beaulieu1}. 

In the proposed system, the symbol that is transmitted by $R$ at the $n^{th}$ time slot, $s_R[n]$, is calculated according to 
\begin{eqnarray}
s_R[n] =  -s_A[n-1] \times s_B[n-1].       
\label{eq:3}
\end{eqnarray} 
Here, $s_R[n]$ is determined based on the joint estimates of $s_A[n-1]$ and $s_B[n-1]$ obtained at $R$. Details of this joint detection process are given in the following subsection.  Note that the relation in \eqref{eq:3}, corresponds to the modulated XOR of  $d_A[n-1]$ and $d_B[n-1]$. 
In this system, the received baseband signal $r_R[n-1]$ is converted to the network coded symbol $s_R[n]$ without individually decoding $s_A[n-1]$ and $s_B[n-1]$. 
%
%

As the relay node broadcasts a delayed version of the information symbols $s_R[n]$,  the nodes $A$ and $B$ transmit $s_A[n]$ and $s_B[n]$, respectively.  The received signal at the end node $i$ is
\begin{eqnarray}
r_i[n] =  h_{iR}s_R[n]+ \kappa_{i} h_{ii}s_i[n] + w_i[n] \; \; \textrm{for}  \;\; i=A,B , 
\label{eq:4}
\end{eqnarray}
where  $\kappa_{i}$ represents the residual SI coefficient and $w_i[n]$ is the AWGN component with the distribution $\mathcal{N}(0, \sigma_i^2)$.  
It can be observed from the system block diagram Fig. \ref{fig:2} that the MA and BC phases of TWRC are combined within a single time slot.  At the $n^{th}$ time slot the end nodes detect the bit of $(n-1)^{th}$ time slot. Hence bits are detected with a fixed delay of one time slot.


\subsection{Error Performance at the Relay Node}

In the proposed FD-PNC, $R$ needs to jointly detect $s_A[n]$ and $s_B[n]$. Hence, $s_A[n]$ and $s_B[n]$ signals cause interference in the MA phase, limiting the system's error performance. This problem can be addressed by the use of joint detection  techniques as proposed in \cite{JD_1,JD_3}. Using joint detection will aid us to increase the robustness against bit errors in presence of MA interference. We use the max-log approximation  which is an effective method for the application of the ML detection rule to reduce the computational complexity \cite{max_log_ap}. The joint detection process starts with expressing the approximated ML decision rule that can be stated as \cite{zang_PNC}
\begin{equation}
(\hat{s}_A, \hat{s}_B \!) = \textrm {arg}  \min_{s_A,s_B\in\mathcal{X}}  \! \left \lvert {r}_R[n]- \!h_{AR}s_A - \!h_{BR}s_B \right \rvert^2,
\label{eq:5}
\end{equation}
where the alphabet is $\mathcal{X}=\pm 1$. We assume that channel coefficients are available at $R$.   After using the ML decision rule in \eqref{eq:5}, we obtain the estimated values of $s_A[n-1]$ and $s_B[n-1]$, hence the $s_R[n]$ is determined through the equation given in \eqref{eq:3}.  Noting that the Gaussian interference model for Rayleigh fading channels provides the exact bit error rate (BER) expression for BPSK modulated symbols (as proven in \cite{Beaulieu1}),  the exact instantaneous BER expressions of $s_A[n]$ and $s_B[n]$ are equal to 
\begin{equation}
P_i^R(e \vert h_{iR})= Q \left(\sqrt{2 \left \vert  h_{iR} \right \vert^2 \gamma_i} \right).
\label{key}
\end{equation}
Since the absolute square of channel coefficients  have exponential distribution, the average BER of the $s_i$ signal at $R$ can be shown to be
\begin{eqnarray}
P_i^R (e) =  E   \left[ P_i^R(e\vert h_{iR})  \right] = \dfrac{1-\sqrt{\alpha_i}}{2},
\label{eq:6} 
\end{eqnarray}
where ${\alpha_i}= E_i /(E_i+\kappa_R^2E_R+ \sigma_i^2)$. Using this definitions, the average exact bit error expression at $R$ can be calculated according to
\begin{eqnarray}
P^R (e)&=& P_A^R(e) (1-P_B^R(e)) +P_B^R(e) (1-P_A^R(e)), \nonumber \\
 &=& \dfrac{1-\sqrt{\alpha_A \alpha_B}} {2}.
\label{eq:7}
\end{eqnarray}
We present the theoretical end-to-end system BER performance in the next section by using \eqref{eq:7}. 
\subsection{End-to-End Error Performance}
Here, we analyze the error performance of the node $A$ and the analysis will be identical for node $B$. 
In order to determine ${s}_R[n]$ at $A$, the ML detection rule is 
\begin{equation}
\hat{s}_R [n] = \textrm {arg} \!\! \min_{{s}_R\in\{\mathcal{X}\}}  \! \left \lvert {r}_A[n]- h_{AR}{s}_R \right \rvert^2,
\label{eq:8}
\end{equation}
where the estimate of ${s}_R[n]$ is denoted by $\hat{s}_R[n]$.  Here we assume that $h_{AR}$ is available at $A$. Following this step, the estimate of  $s_B[n-1]$ (denoted by $\hat{s}_B[n-1]$), is calculated according to
\begin{equation}
\hat{s}_B[n-1]= -s_A[n-1] \times \hat{s}_R[n],
\label{eq:9}
\end{equation}
since $s_A[n-1]$ is known $a \; priori$ at $A$. If $R$ correctly determines $s_R[n]$, the instantaneous error probability at $A$ is equal to
\begin{equation}
P_B^A (e\vert h_{iR})= Q\left(\sqrt{ 2 \left \vert  h_{AR} \right \vert ^2 \gamma_R } \right),
\label{}
\end{equation}
where $\gamma_R=E_R/ \left( \sigma_A^2 + \kappa_A^2E_A\right)$. The error rate at node $A$ is a function of $P^R(e)$  and  the average BER is equal to $P_B^A(e)=E[P_B^A (e\vert h_{iR})]$. Here two events ($P^R(e)$ and $P_B^A(e)$) are statistically independent due to physical separation of nodes. When $s_R[n]$ is transmitted in error by $R$, and $A$ incorrectly detects the erroneously transmitted bit,  $s_R[n]$ signal is obtained correctly. Hence the average error probability of the end-to-end system is equal to
 \begin{eqnarray}
 P^A (e) \! \!\!&=& \!\!\!\! P_{}^R(e)(1-P_B^A(e)) +(1- P_E^R(e))P_B^A(e), \nonumber \\
   &    =&  \dfrac{1-\sqrt{\alpha_A \alpha_B \alpha_R}}{2}
   \label{eq:11}
 \end{eqnarray}
where $\alpha_R=E_R/ (E_R+ \kappa_A^2E_i+ \sigma_A^2)$. 

As the signal to noise ratio of communication channels increase, the effect of SI becomes more dominant, generating error-rate floors. The error-rate floor at $A$,  $\varGamma_A$, can be obtained as 
\begin{eqnarray}
\varGamma_A&=&\lim_{\sigma_R^2,\sigma_i^2\to0 } \left\{P^A(e) \right\}, \nonumber  \\ \!\!\!\!\!\!\!\!\!\!&=& \frac{1}{2}- \frac{1}{2(1+\kappa_R^2)\sqrt{1+\kappa_A^2}} ,
\label{eq:12}
\end{eqnarray}
for the case of unit energy transmissions.
\section{Numerical and Simulation Results}

In this section, theoretical and matching simulation results are given for the FD-PNC system in presence of SI for all nodes. The results are presented for relay and end-to-end BER performances given in \eqref{eq:7} and \eqref{eq:11}, respectively.  We use BPSK modulated signals, distribution of all channels are Rayleigh fading and required channel coefficients are assumed to be known at all nodes. The analyses can be extended to higher order modulations through similar steps.

\begin{figure}[t]
\centering
\includegraphics[width=3.55in, keepaspectratio=true] {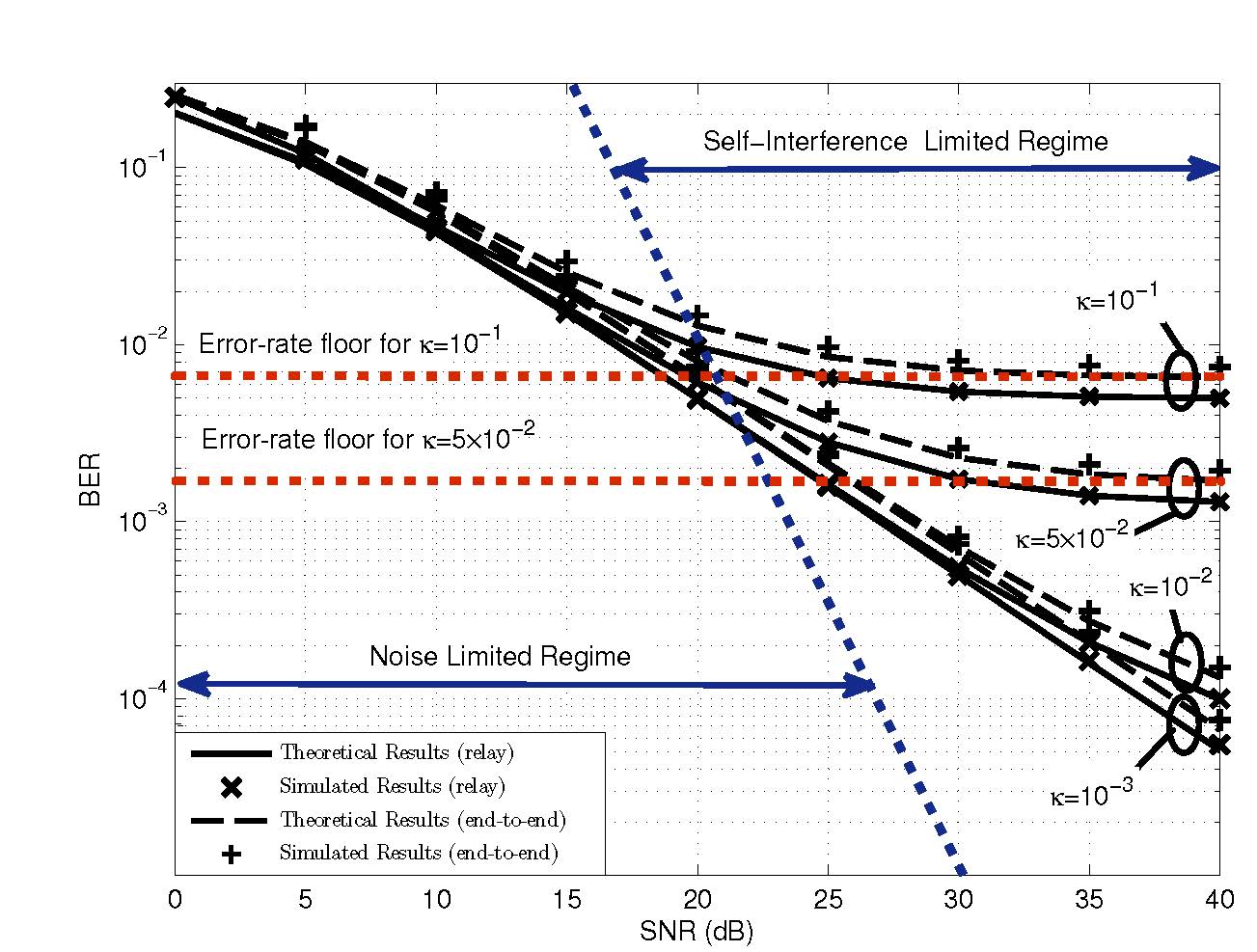}
\caption{Relay node and end-to-end bit error rate performances of the proposed FD-PNC system for various $\kappa$ values ($\kappa_R=\kappa_A=\kappa$). The approximate boundary between the self-interference and noise dominated operating regimes are also shown.} 
\label{fig:3}
\end{figure}

BER curves for the end-to-end system and the relay nodes are shown in  Fig \ref{fig:3} for various residual SI coefficients for nodes $R$ and $A$. Note  that theoretical results match simulation results, confirming the validity of the Gaussian interference model \cite{JD_1,Beaulieu1}. In this figure we can identify two operating regimes, the noise limited regime ($\sigma_i^2 > 2\kappa ^2$ for unit energy transmissions) and the SI limited regime ($\sigma_i^2 \leq 2\kappa ^2$ ). 
The error-rate floors of the system depending on the residual SI component given in \eqref{eq:12} are also highlighted in the figure. 

\begin{figure}[t]
\centering
\includegraphics[width=3.5in, keepaspectratio=true] {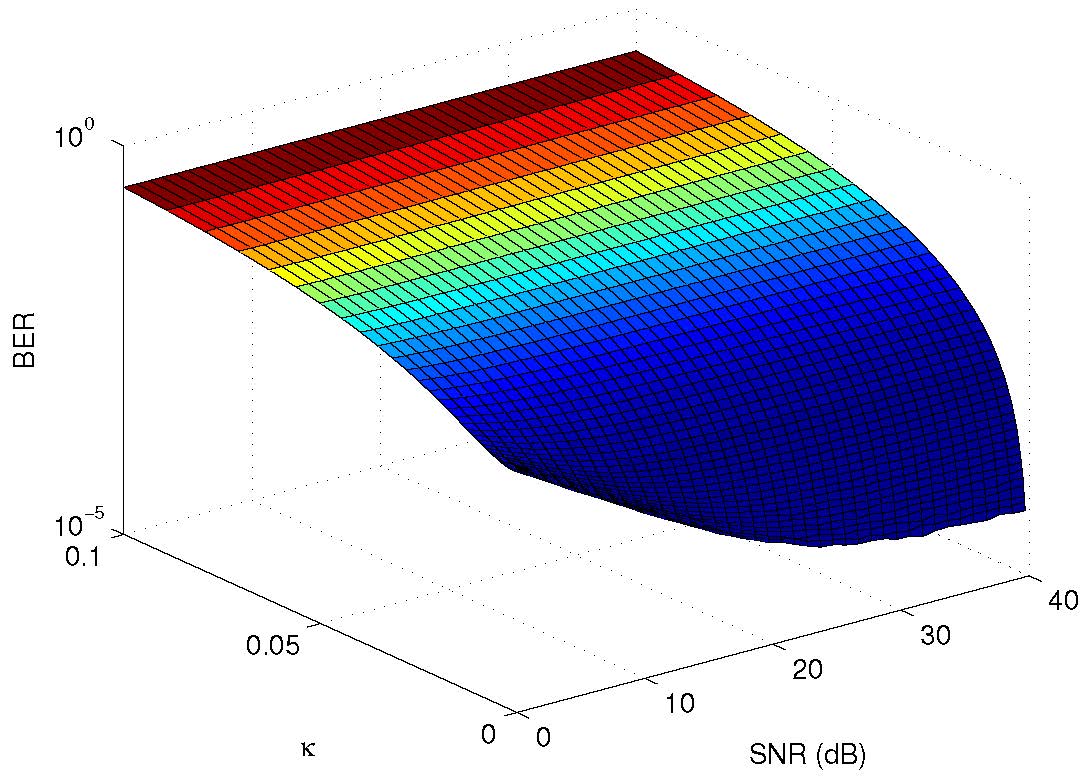}
\caption{End-to-end bit error rate of the proposed FD-PNC system for different $\kappa$ and $SNR$ values where $\kappa_R=\kappa_A=\kappa$. Acceptable operating points can be achieved with high $SNR$ values even when $\kappa \neq 0$.  }
\label{fig:4}
\end{figure}

As noted throughout our analyses, the effects caused by the residual SI coefficients can not be ignored. Fig. \ref{fig:4}   demonstrates the  performance degradation due to imperfect implementation of transmit and receive antenna chains of the FD nodes. From this figure we can see that for the non-ideal case of SI cancellation $\kappa_A=\kappa_R = 2 \times 10^{-4}$ (which equivalently models the 74 dB SI suppression implemented  with consumer of the shelf equipments in \cite{duarte}) we can obtain a BER of  $10^{-4}$   for $SNR \approx 38$ dB, verifying the practicality of the proposed FD-PNC.

\begin{figure}[t!]
\centering
\includegraphics[width=3.5in, keepaspectratio=true] {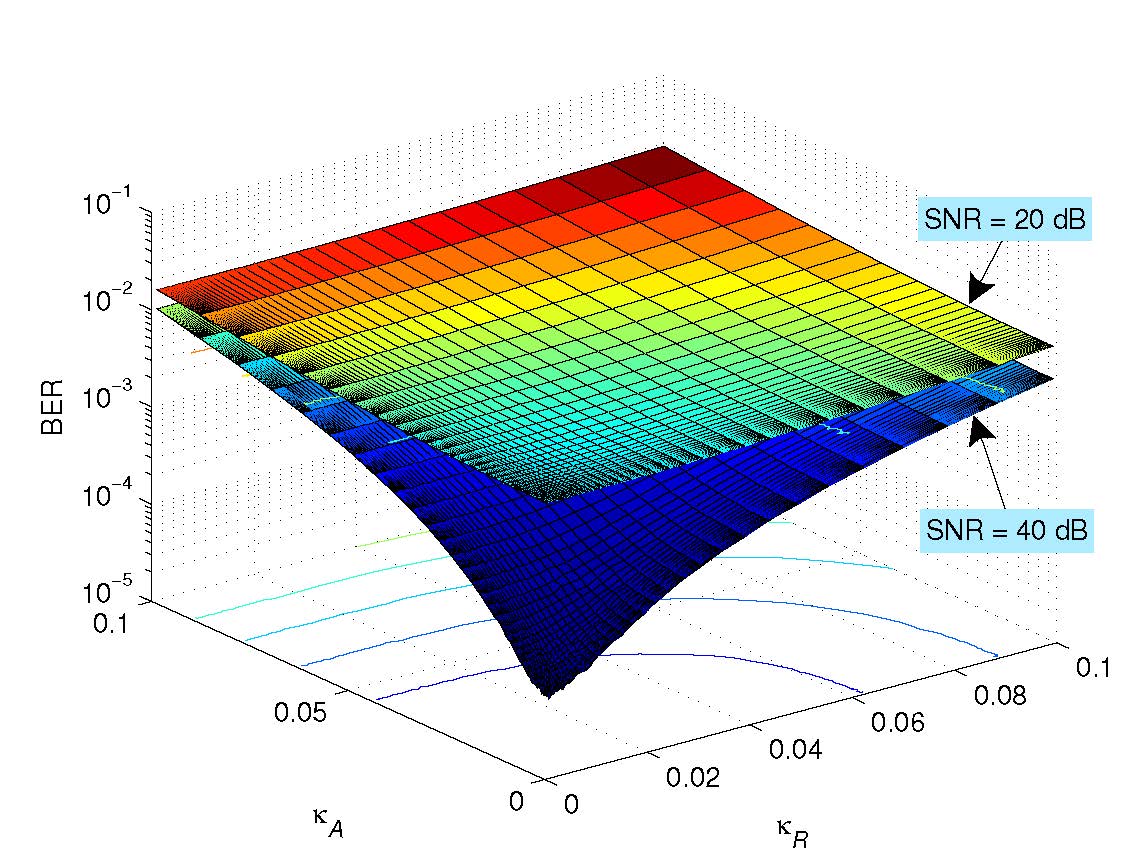}
\caption{End-to-end bit error rate performances of the proposed FD-PNC system for different $\kappa_R$ and $\kappa_A$ values. It can be observed that the residual self-interference in the relay node deteriorates the performance more that the residual self-interference at the end nodes. }
\label{fig:5}
\end{figure}

Finally noting that the residual SI at the relay node has more deteriorating effects on the end-to-end BER,  in Fig. \ref{fig:5}  we quantify the impact of $\kappa_A$ and $\kappa_R $ on the BER performance for $SNR = 20$ dB and $SNR = 40$ dB. 
From this figure  we can observe that the error rates can be  improved if  better performing FD nodes are selected as the relay nodes, rather than the end nodes.  

\section{Conclusion}
In this paper, we proposed a physical layer network code for TWRC  with FD  nodes, where each node is equipped with a single transmit and receive antenna. The proposed system decreased the total transmission time of two time slots of classical PNC systems to a single time slot, executing the MA and BA phases concurrently. 
 The theoretical BER expressions in case of non-ideal SI cancellation over Rayleigh fading channels are derived for relay and end nodes. Monte Carlo simulations are used to verify the theoretical results.
\bibliography{ref_genel} 

\bibliographystyle{IEEEtran}

\end{document}